\documentclass[useAMS,usenatbib]{mn2e} \usepackage[T1]{fontenc}
\usepackage{ae,aecompl} \usepackage{times}

\usepackage{graphicx}
\usepackage[varg]{txfonts}
   \title[e-MERLIN resolves Betelgeuse at $\lambda$ 5 cm] {e-MERLIN
resolves Betelgeuse at $\lambda$ 5 cm: 
hotspots at 5 R$_{\star}$}

   \author[A. M. S. Richards et al.]{A.~M.~S.~Richards$^{1}$,
 R.~J.~Davis$^{1}$, L. Decin$^{2}$,
S. Etoka$^{1,3}$, G.~M.~Harper$^{4}$ J.~J. Lim$^{5,6}$, \newauthor
S.~T. Garrington$^{1}$, M~.D. Gray$^{1}$, I. McDonald$^{1}$,
E.~O'Gorman$^{4}$, M. Wittkowski$^{7}$\\ $^1$JBCA, Dept. Physics and
Astronomy, University of Manchester, UK; $^2$Instituut voor
Sterrenkunde, Katholieke Universiteit Leuven, Belgium\\ $^3$Hamburger
Sternwarte, Germany; $^4$School of Physics, Trinity College, Dublin 2,
Ireland; $^5$Department of Physics, University of Hong Kong;\\
$^6$Institute of Astronomy and Astrophysics, Academia Sinica, Taipei,
Taiwan; $^7$ESO, Garching bei M\"{u}nchen, Germany.  } 
\date{Accepted MNRAS March 12, 2013}

\pagerange{\pageref{firstpage}--\pageref{lastpage}} \pubyear{3012}

\def\LaTeX{L\kern-.36em\raise.3ex\hbox{a}\kern-.15em
T\kern-.1667em\lower.7ex\hbox{E}\kern-.125emX}

\begin{document}
\label{firstpage}

\maketitle

\begin{abstract} Convection, pulsation and magnetic fields have all
  been suggested as mechanisms for the transport of mass and energy
  from the optical photosphere of red supergiants, out to the region
  where the stellar wind is launched.  We imaged the red supergiant
  Betelgeuse at $0.06-0.18$ arcsec resolution, using e-MERLIN at
  5.5--6.0 GHz, with a sensitivity of $\sim$10 $\mu$Jy beam$^{-1}$.  Most
  of the radio emission comes from within an ellipse
  (0.235$\times$0.218) arcsec$^2$ ($\sim$5$\times$ the optical radius),
  with a flux density of 1.62 mJy, giving an average brightness
  temperature $\sim$1250 K.  This radio photosphere contains two
  hotspots of 0.53 and 0.79 mJy beam$^{-1}$, separated by 90
  milli-arcsec, with brightness temperatures $5400\pm600$ K and
  $3800\pm500$ K.  Similar hotspots, at more than double the
    distance from the photosphere of those seen in any other regime,
  were detected by the less-sensitive `old' MERLIN in 1992, 1995 and
  1996 and many exceed the photospheric temperature of 3600 K.
  Such brightness temperatures are high enough to emanate from pockets
  of chromospheric plasma.  Other possibilities include local shock
  heating, the convective dredge-up of hot material or exceptionally
  cool, low density regions, transparent down to the  hottest
    layer at $\sim$40 milliarcsec radius.  We also detect an arc
  0.2--0.3 arcsec to the SW, brightness temperature $\sim 150$ K, in a
  similar direction to extensions seen on both smaller and larger
  scales in the infra-red and in CO at mm wavelengths. These
  preliminary results will be followed by further e-MERLIN, VLA and
  ALMA observations to help resolve the problem of mass elevation from
  1 to 10 $R_{\star}$ in red supergiants.
\end{abstract}

%

\section{Introduction}
\label{sec:intro} Betelgeuse ($\alpha$ Ori), type M2Iab, is the
closest red supergiant, with an initial mass of 15--20 M$_{\odot}$
\citep{Dolan08}.  The asymmetric S1 and S2 CO outflows have average
velocities and radius limits of 9.8 km s$^{-1}$, 4 arcsec and 14.3 km
s$^{-1}$, $\sim$17 arcsec, respectively \citep{OGorman12}.  The
mass-loss rate $\dot{M}$ of Betelgeuse is $\sim$$10^{-6}$ M$_{\odot}$
yr $^{-1}$, estimated by \citet{LeBertre12} from a 0.24-pc HI shell.
Its proper motion is towards the NE and a bow shock was identified in
the infra-red (IR) \citep{Noriega-Crespo97} and HI and mapped using
\emph{Herschel} \citep{Decin12}, which shows arcs 6--7 arcmin NE of
the star. We adopt the 2.2 $\mu$m photospheric
radius $\sim$22.5 milli-arcsec (mas) (\citealt{Dyck96};
\citealt{Perrin04}) as $R_{\star}$, at a distance of $\sim$197 pc
\citep{Harper08}.

 Mass loss from cool,
evolved stars (e.g. VX Sgr) which form copious dust at $\sim$5--10
$R_{\star}$, is thought to be initiated by pulsations, followed by
radiation pressure on dust accelerating the wind to escape velocity
\citep{Bowen88}. However, the requirements for this mechanism do not
seem to be present in early M-type stars like Betelgeuse.  It shows
weaker optical variability \citep{Percy96} than the RSG which form
more dust.  \citet{Verhoelst06} found signs of metal oxide nucleation
at $\sim$1.5 $R_{\star}$ but silicates have only been found at radii
$\ga$500 mas (e.g. \citealt{Tatebe07}; \citealt{Skinner97}).
\citet{Dupree87} measured periodicities in optical and ultraviolet
fluxes probably tracing pulsatational shock waves which could help to
initiate mass outflow.  Other potential mass-loss mechanisms include
localised events related to convection (e.g. as modelled by
\citealt{Chiavassa10}).  \citet{Auriere10} measured a longitudinal
magnetic field of 0.5--1.5 G, leading to models for magnetic promotion
of mass loss \citep{Thirumalai12}.  See the Proceedings of the
Betelgeuse 2012 workshop, eds. Kervella et al., in
prep., for more comprehensive reviews of  mass loss studies.

VLA observations by \citet{Lim98} at 43, 22, 15, 8 and 5 GHz showed
the measured radius increasing from 2--7 $R_{\star}$. The opacity
increases with decreasing frequency, so lower frequency images sample
layers at larger radii, found to be cooler. Over this range of
$R_{\star}$, the radio brightness temperature $T_{\mathrm b}$ falls
from $\sim$$3540-1370$ K and the photosphere becomes cool enough for
dust to begin to form.  However, \citet{Uitenbroek98} detected UV
continuum and lines formed at $\ga$5000 K, at radii of up to 2.8 and 6
$R_{\star}$, respectively.   Absorption lines reveal a rotation speed
of 5 km s$^{-1}$ and they suggest that a bright spot in the SW defines
the orientation of the rotational axis, at a position angle of
55$^{\circ}$, giving a deprojected rotation period of 25 yr at
14$^{\circ}$ yr$^{-1}$.  The chromospheric tracer Mg{\small {II}} shows an
expansion velocity of 10 km s$^{-1}$, detected out to almost 9
$R_{\star}$ \citep{Gilliland96}.  Asymmetric absorption lines such as
H$\alpha$  (\citealt{Weymann62};
\citealt{Bagnulo03}) suggest outflow of chromospheric material at 5--7
km s$^{-1}$.

\citet{Lim98} deduced that the chromospheric gas must be at least a
thousand times less abundant than the material dominating radio
emission, ruling out global heating as a mass-loss mechanism, leaving
convection as a plausible alternative.  \citet{Harper06} studied
multi-epoch spatially resolved HST spectra and, by measuring the
electron density and comparing it to 22-GHz observations, confirmed
the small filling factor. Chromospheric emission lines suggested
rotation about an axis at a position angle of 65$^{\circ}$ at a
velocity not exceeding $\sim5$ km s$^{-1}$, suggesting that
chromospheric material is not in solid-body rotation at $R>75$ mas.

We observed Betelgeuse using e-MERLIN at 5.5--6.0 GHz
($\lambda$$\approx$5.2 cm), providing $\sim4$ resolution elements
across the stellar diameter.  These are the first well-resolved images
at cm wavelengths, investigating whether the irregularities seen at 43
GHz ($\lambda 7$ mm) persist at larger radii.  We describe the
observations in Section~\ref{sec:obs} and results in
Section~\ref{sec:results}, discussing these in
Section~\ref{sec:discussion}. We suggest future work in
Section~\ref{sec:future} and conclude with a summary in
Section~\ref{sec:summary}.

\section{Observations and data reduction}
\label{sec:obs}
\subsection{2012 e-MERLIN observations} Betelgeuse was observed in the
first semester of e-MERLIN open time on 13--15 July 2012, with a
bandwidth of 0.512 GHz centred on 5.75 GHz. Seven antennas were used
including the 75-m Lovell telescope providing baselines from 11--217
km (90--3910 k$\lambda$). The data were processed in dual
polarization, using 4$\times$128 MHz spectral windows, each divided
into 64 2-MHz channels.  The point-like QSO 0551+0829, separation
$~1.\!^{\circ}5$, was used as the phase reference on a cycle of 7:3
min. OQ208 was used as the bandpass and flux density calibrator.
Calibration is summarized in Appendix A. The flux scale is accurate to
$\sim$10\% and the astrometric uncertainty arising from phase
referencing is 16 mas.  The calibrated and edited Betelgeuse data
comprised 4--8 hr per antenna, spread over 10.5 hr, with an average
bandwidth of 400 MHz.

We made two maps after self-calibration, using different weighting
schemes (see Appendix A). We maximised sensitivity using a
3500-k$\lambda$ taper and a circular 180-mas (FWHM) restoring beam.
This produced a noise $\sigma_{\mathrm {rms}}$ 0.027 mJy beam$^{-1}$.
We maximised resolution by applying partial uniform weighting
({\small {ROBUST 0.75}}) and using a restoring beam of ($80\times60$)
mas$^2$ at position angle (PA) 143$^{\circ}$ (close to the natural
beam shape), giving $\sigma_{\mathrm {rms}}$ 0.009 mJy
beam$^{-1}$. Despite the lower noise per beam, the minimum brightness
temperature sensitivity limit is 225\% higher with this weighting.  In
order to ensure artifacts were not introduced, we made images with
completely natural weighting (used for astrometry) and with the above
weightings, prior to self-calibration. The high-resolution dirty map
has $\sigma_{\mathrm{rms}}$ of 0.03 mJy beam$^{-1}$, with peaks of
0.81 and 0.63 mJy beam$^{-1}$, similar to the features described in
Section~\ref{sec:hotspot}.

\subsection{Archival data} We retrieved MERLIN Betelgeuse data
observed at 4.994 GHz on 1995-06-24 and 1996-11-03 \citep{Morris01}
using 16 MHz bandwidth, and VLA data observed at 4.885 GHz on
1996-10-21 using 100 MHz bandwidth.  These data sets were reduced
using standard techniques (\citealt{Diamond03}, \citealt{Greisen94}).
The MERLIN data were imaged with natural weighting (beam size
($85\times55$) mas$^2$) and no cleaning, giving $\sigma_{\mathrm
{rms}}$ of 0.15 and 0.12 mJy beam$^{-1}$ in 1995 and 1996,
respectively.  The 1996 MERLIN and VLA data were combined to give a
resolution of 200 mas and $\sigma_{\mathrm {rms}}$ of 0.058 mJy
beam$^{-1}$; the VLA-only image had 400-mas resolution.

\section{Results}
\label{sec:results}

   \begin{figure} \centering
   \includegraphics[width=\hsize]{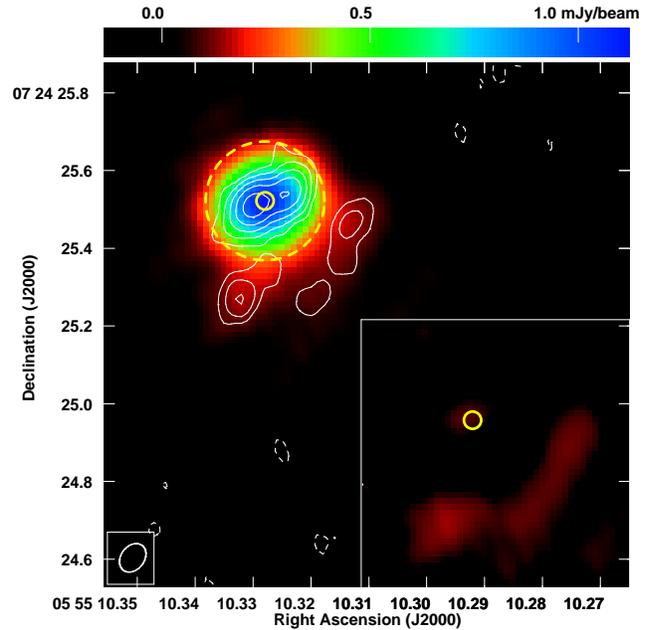}
   \caption{The main panel shows the 2012 e-MERLIN colour image of
Betelgeuse optimised for sensitivity to extended structure, using a
beam size of 180 mas, $\sigma_{\mathrm {rms}}$ = 0.027 mJy
beam$^{-1}$.  The contours of the image optimised for resolution are
overlaid, with the beam size at lower left, the noise $\sigma_{\mathrm
{rms}}$ = 0.009 mJy beam$^{-1}$, and contour levels at (--1, 1, 2, 4,
8, 16) $\times$ 0.027 mJy beam$^{-1}$. The small, solid yellow circle
is the size of the 45-mas diameter photosphere, centred on the
low-resolution peak, and the large, dashed circle shows the 310-mas,
5-GHz disc \citep{Lim98}.  The insert shows the SW arc after
subtracting the central star, on the same scale.  }
         \label{AORI2012_ALL.CPS}
   \end{figure}

\subsection{Extended structure seen in 2012}
\label{sec:extnd}

Figure~\ref{AORI2012_ALL.CPS} shows the colour image made with optimum
sensitivity to extended structure, using a 180-mas restoring beam. The
peak is 1.116 mJy beam$^{-1}$.  The 3$\sigma_{\mathrm {rms}}$ boundary
(0.081 mJy beam$^{-1}$ at this resolution), represented by the limit
of red shading, has a quite irregular outline, with a maximum extent
of 550 mas approximately N--S.   The
shortest, radial separation between the peak and the 3$\sigma_{\mathrm
{rms}}$ limit is about 190 mas, to the NNE.  We fitted a 2D elliptical
Gaussian component to this image, yielding a total flux density
$1.619\pm0.057$ mJy at position 05:55:10.3274 +07:24:25.514, with a
noise-based uncertainty of 4 mas.  The component had major and minor
axes ($\theta_{\mathrm {maj}}$, $\theta_{\mathrm {min}}$) of
(235$\times$218) mas$^2$, uncertainty $\sigma_{\theta}$ = 6 mas, PA
$115\pm9^{\circ}$ (significantly different from the beam PA of
143$^{\circ}$).

\begin{figure} \centering
   \includegraphics[width=\hsize]{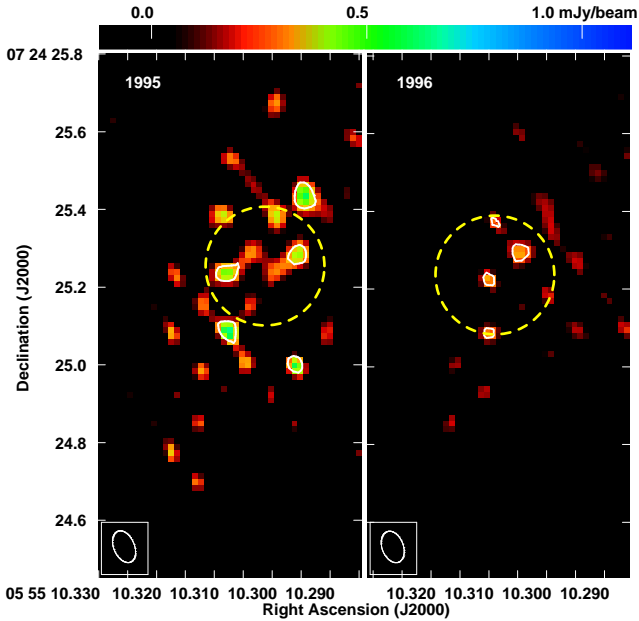}
   \caption{MERLIN images of Betelgeuse observed at 4.994 GHz in 1995
and 1996, contours at 0.60 and 0.48 mJy beam$^{-1}$ (4
$\sigma_{\mathrm {rms}}$), respectively, using the same angular scale
as Fig.~\ref{AORI2012_ALL.CPS}. The beam size is (85$\times$55)
mas$^2$. The large, dashed circle shows the 310-mas, 5-GHz disc
\citep{Lim98}.  }
         \label{MERLIN95-96.PS}
   \end{figure}

The brightness temperature of a component of flux density $S$, area
$A$, is given in Kelvin by
\begin{equation} T_{\mathrm b} = 15400 \left(\frac{S}{\mathrm
{mJy}}\right) \left(\frac{\lambda}{\mathrm
m}\right)^2\left(\frac{A}{\mathrm {arcsec}^2}\right)^{-1} .
\end{equation} 
We observed at wavelength $\lambda$ = 0.052 m.  The
area of a Gaussian component is given by $A=\pi(\theta_{\mathrm {maj}}
\theta_{\mathrm {min}})/(4\ln2)$. The main component thus has
$T_{\mathrm b} = 1170\pm135$ K.  There is a point-like residual of
0.082 mJy beam$^{-1}$ (only just over 3$\sigma_{\mathrm {rms}}$) at
the position of the peak, contributing another $\sim$100 K, giving
$T_{\mathrm b} = 1250\pm150$ K within a radius of $\sim$5 $R_{\star}$.
This is close to the temperature and the size (dashed circle in
Figs.~\ref{AORI2012_ALL.CPS} to~\ref{MV_96.CPS}) measured by
\citet{Lim98} by fitting a uniform disc to the 5-GHz VLA visibilities,
showing the stability of these atmospheric conditions over 16 yr.

 The residuals after subtracting the central component are shown in
the insert of Fig.~\ref{AORI2012_ALL.CPS}.  There is a substantial arc
in the SW quadrant, with a maximum extent above 3$\sigma_{\mathrm
{rms}}$ of 510 mas with a total flux density 0.088$\pm0.019$ mJy in
0.0249 arcsec$^2$.  It has an irregular outline but is approximately
100 mas wide, at a radius between $\sim$175--275 mas from the central
peak.  The SW arc has an average $T_{\mathrm b} = 150\pm40$ K.

\subsection{Position and flux density}
\label{sec:pos}
Our first map of Betelgeuse (phase-reference solutions only) had a
peak of 0.78$\pm$0.063 mJy beam$^{-1}$ at 05:55:10.3267 +07:24:25.525.
The noise-based position uncertainty is 9 mas and the total
astrometric uncertainty (mostly due to the $1.\!^{\circ}5$ separation
from the phase reference source) is 20 mas.  \citet{Harper08} used
\emph{Hipparcos} and multi-epoch VLA data to solve for position,
proper motion and parallax, up to epoch 2004.829.  The preferred
solution (their number 5) predicts a position of 05:55:10.3250
+07:24:25.536, uncertainty (5, 6) mas at our epoch, 2012.536. This is
(25, 10) mas from our position, close to the combined uncertainties.

The total flux density of 1.62--1.71 mJy (depending on whether the SW
arc is enclosed) measured at 5.75 GHz, is equivalent to
(1.3--1.4)$\pm0.2$ mJy at 4.85 GHz, using a spectral index of 1.32
\citep{Newell82}, allowing for the uncertainty in this
extrapolation. Observations at 4-8--4.9 GHz showed a decline from 2 to
1.2 mJy during 1981 to 2002, fluctuating within the range 0.9--1.3 mJy
up to 2004 (\citealt{Newell82}, \citealt{Skinner97}, \citealt{Lim98};
Harper \& Brown private communication). 

\subsection{Hotspots seen in 2012}
\label{sec:hotspot} The contours in Fig.~\ref{AORI2012_ALL.CPS} show
the image optimised for resolution, using a ($80\times60$) mas$^2$
restoring beam. Two individually-unresolved peaks appear, with flux
densities and positions of 0.706 mJy beam$^{-1}$ at 05:55:10.3299
+07:24:25.510 and 0.489 mJy beam$^{-1}$ at 05:55:10.3241
+07:24:25.541, respectively (measured by fitting two Gaussian
components simultaneously).  $\sigma_{\mathrm {rms}}$=0.0095 mJy
beam$^{-1}$ and the noise-based position uncertainties are 1 mas for
the brighter peak and 2 mas for the fainter one.  The hotspots are
separated by $90\pm10$ mas, at a PA of $(110\pm10)^{\circ}$. This is
similar to the orientation of the major axis of the Gaussian component
fitted to the emission imaged at 180-mas resolution
(Section~\ref{sec:extnd}) and significantly different from the beam
PA.  The peaks are separated by $>20\sigma$ in both position and flux
density.  Their brightness temperatures are $T_{\mathrm b} =
5400\pm600$ K and $T_{\mathrm b} = 3800\pm500$ K, uncertainties
dominated by the overall flux scale uncertainty.

   \begin{figure} \centering
   \includegraphics[width=\hsize]{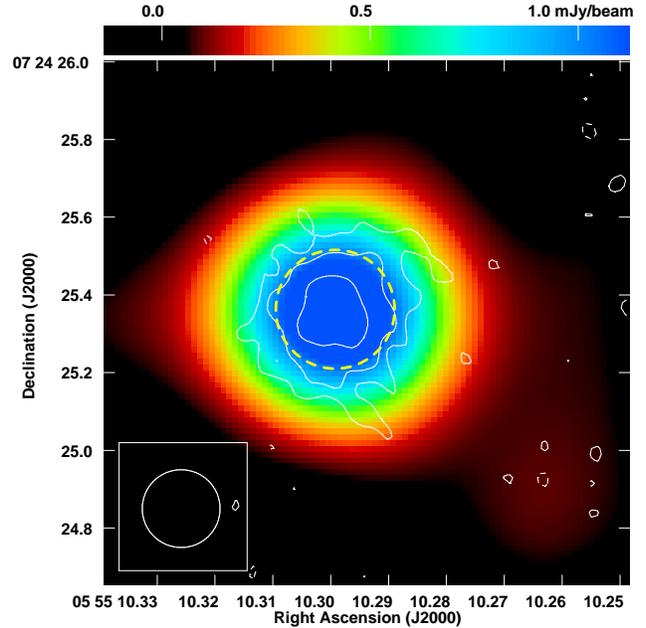}
   \caption{1996 combined MERLIN+VLA image of Betelgeuse (contours at
(--1, 1, 2, 4) $\times$ 0.17 mJy beam$^{-1}$), beam size 200 mas,
overlaid on the VLA-only image observed at 4.9 GHz, beam size 400 mas.
The angular scale is the same as Fig.~\ref{AORI2012_ALL.CPS}.  The
large, dashed circle shows the 310-mas, 5-GHz disc \citep{Lim98}. }
         \label{MV_96.CPS}
   \end{figure}

\subsection{1995 -- 1996 MERLIN and VLA results}

Figure~\ref{MERLIN95-96.PS} shows that, at the lower sensitivity of
the `old' MERLIN archive data, just a few hotspots were detected at
the position of Betelgeuse.  In 1995, 5 spots brighter than 0.60 mJy
beam$^{-1}$ ($4 \sigma_{\mathrm {rms}}$) were detected, the brightest
at 0.78 mJy beam$^{-1}$.  In 1996, 4 spots brighter than 0.48 mJy
beam$^{-1}$ ($4 \sigma_{\mathrm {rms}}$) were detected, the brightest
at 0.56 mJy beam$^{-1}$.  The formal position uncertainties are
$\sim$40 mas but may be higher due to calibration errors.  The
shortest baseline (11 km) allows scales as large as 0.5 arcsec to be
imaged, and the sensitivity of the 1995 and 1996 images corresponds
approximately to the highest contour in Fig.~\ref{AORI2012_ALL.CPS},
so the non-detection of most of the stellar disc was due to lack of
sensitivity rather than missing spacings.  The peak brightness
temperatures (at $\lambda=0.061$ m) were $T_{\mathrm b} = 8000\pm2000$
K in 1995 and $T_{\mathrm b} = 6000\pm2000$ K in 1996. The
uncertainties are too high to draw any firm conclusions from the
distribution of the spots.

The contours in Fig.~\ref{MV_96.CPS} show the 1996 combined MERLIN+VLA
image, containing $1.69\pm0.17$ mJy in a Gaussian component with
($\theta_{\mathrm {maj}}$, $\theta_{\mathrm {min}}$) of
($289\times256$) mas$^2$, $\sigma_{\theta}$ 17 mas, PA
$(150\pm30)^{\circ}$. This gives $T_{\mathrm b} = 1200\pm200$ K.  The
colour image shows the 1996 VLA-only data, which were part of a
multi-frequency dataset analysed by \citet{Lim98} and
\citet{Harper03}, who measured a flux density of $1.77\pm0.09$ mJy.
The VLA image contains a faint feature 0.6 arcsec due SW of the peak
(over twice the distance to the e-MERLIN arc), at 0.095 mJy
beam$^{-1}$, just over 3$\sigma_{\mathrm {rms}}$.

\section{Discussion}
\label{sec:discussion}
\subsection{SW Arc}

The SW arc is at a radius of $\sim$250 mas, where, according to the
semi-empirical model of \citet{Harper01}, the electron temperature
$T_{\mathrm e}$ is $\sim$900 K and the hydrogen number density 
$n_{\mathrm H}$$\sim$5$\times$$10^{13}$ m$^{-3}$.  This gives an optical depth $\sim$0.18
for the observed $T_{\mathrm b} \sim$150 K.  Assuming that the arc is
as deep as it is wide, we approximate the region producing most of the
emission by a spheroid with semi-major axes of 10, 10 and 25 au,
giving a hydrogen mass of $\sim1.8\times10^{-6}$ M$_{\odot}$,
suggesting that one such clump could be formed every couple of years.

The arc is about 100 mas outside the 6-GHz photospheric radius, in a
similar direction to the CN plume seen by \citet{Kervella09} at $\la6$
$R_{\star}$ and the faint SW extension at $\sim0.6$ arcsec seen at 5
GHz in 1996. A clump of CO emission was seen in 2007 by
\citet{OGorman12} just outside the S1 shell, 5 arcsec SW of the
centre, suggesting a wind travel time of $\sim$500 yr at 9 km
s$^{-1}$.  These observations suggest repeated or continuous
ejections, rather than proper motion of a single clump. This is also
aligned with the SW pole of the rotational axis identified by
\citet{Uitenbroek98}. The direction could be coincidence, since
observations at 7.76--19.5 $\mu$m in Nov 2010 \citep{Kervella11} show
clumps outside the photosphere at 0.8--1.7 arcsec around more than a
semi-circle, between NE and S.  A comparison with the direction of the
bow shock seen 6--7 arcmin to the NE \citep{Decin12} is tempting, but
structures on sub-arcmin scales are deep within the astropause and
should share the proper motion of the star.

\subsection{Hotspots} The 5--6 GHz hotspots, measured at 55--85 mas
resolution, have $T_{\mathrm{b}}$ formally above the peak photospheric
temperature of 3600 K \citep{Dyck96}, similar to the chromospheric
temperature, which could reach 8000 K \citep{Harper06}.  We used
conservative uncertainties, dominated by the maximum uncertainty in
the flux scale, although the good agreement with VLA results for the
whole-disc flux density suggests our uncertainties may be
overestimated.  One or more hotspots are significantly hotter than the
photosphere in 2012, 1996 and 1995 and also in the 1992 MERLIN
observations imaged at full resolution \citep{Morris01}.  Note that,
at the sensitivity of old MERLIN, no emission at all at $T_{\mathrm b}
\la4500-6000$ K (depending on epoch) would have been detected.

The radio photosphere major axis is at $110-115^{\circ}$ in the
fully-calibrated 2012 data, compared with (67$\pm$7)$^{\circ}$ at
43-GHz in 1998 \citep{Lim98}.  The orientation of the optical disc and
hot spots also varies; the $\sim20$ mas axis joining the two hotspots
seen by \citet{Haubois09} is approximately NE--SW.  Four sets of IR
observations over 24 months by \citet{Tuthill97} showed 2--3 hotspots
at a variety of position angles within the inner $\sim100$ mas.

\citet{Harper01p} showed that increasing the local $T_{\mathrm e}$
2--3-fold enhances radio emission much more effectively than
increasing the gas density by the same factor.  There are several
possible origins for hotspots:
\begin{enumerate}
\item Levitation of gas from the inner photosphere, e.g. by
convection, although such gas would cool radiatively and by expansion.
\item Exceptionally cool, small patches in the outer layers,
transparent enough to expose hotter, inner layers. The separation of
the 2012 hotspots is $90\pm10$ mas, similar to the 80-mas diameter of
the hottest layer  (where the greatest chromospheric contribution to
emission is greatest, \citealt{Harper06}, using 197 pc distance).
\item Shock heating due to pulsations and/or convection;
\item Chromospheric patches at $\sim$5 $R_{\star}$ heated by acoustic and/or magnetic
processes.
\end{enumerate}

Options (iii) and/or (iv) seem most likely to produce radio hot spots
$\ga$4000 K. \citet{Ireland11} modelled pulsation in Miras out to $5
R_{\star}$, finding average effective temperatures $\le$3800 K. Large-scale up- and down-flows are seen in
chromospheric lines, including directional reversals at up to
$\sim$3$R_{\star}$, measured by \citet{Lobel01}.  \citet{Ohnaka11}
measured up- and down-draughts of CO, within $\sim$2.5$R_{\star}$, at
$\la$30 km s$^{-1}$, with different velocities occurring
simultaneously on different sides of the star.  Such speeds are
sufficient to supply shock heating even if it is not clear how
convection could operate at $\gg$2.5$R_{\star}$.  \citet{Wittkowski11}
made spectro-interferometric observations, around 2 $\mu$m, of
continuum and clumpy molecular layers around AGB stars, which
suggested pulsation- and shock-induced chaotic motion, not requiring
convection. Most radio emission is non-chromospheric
(Section~\ref{sec:intro}), but the hotspots have $T_{\mathrm b}
\la$8000 K, in the chromospheric range. \citet{Harper06} suggest that
chromospheric gas is confined within magnetic fields; dissipation of
magnetic energy may continue to heat the gas even at high elevations.

\section{Further high-resolution studies}
\label{sec:future} We have shown that both cool, extended
radio-continuum emission and hot starspots on Betelgeuse can be
resolved at 5.5--6 GHz.  Multi-epoch, multi-frequency follow-up is
needed to establish the emission mechanisms and role in mass loss.  In
full operations, e-MERLIN will reach 0.002 mJy beam$^{-1}$ in two full
tracks at 5--7 GHz. Proper motions of 5 km s$^{-1}$ could be measured
in 4 months for a spot 0.3 mJy brighter than its surroundings,
revealing whether the hotspots rotate as predicted for the
chromosphere.  The variability timescale will constrain the underlying
mechanism. Modelling by  \citet{Harper01p} shows that C{\small {II}} recombination to C{\small {I}} takes $\sim2$
months at $n_{\mathrm H}$$\sim$4$\times$$10^{14}$ m$^{-3}$, around 5 $R_{\star}$, or longer at higher radius, and chemical
changes have similar timescales, but mechanisms
involving bulk up-/down-draughts alone would take about a year to
produce comparable changes.

Matching resolution at 50 mas can be achieved by the EVN, e-MERLIN and
the Karl G. Jansky VLA (alone or in combination as required) at
1.4--50 GHz, and eventually by ALMA at higher frequencies.  Finer
resolution ($\le20$ mas) is possible at 22 GHz and some higher
frequencies, suited to the smaller observed size of the radio
photosphere.  This will allow us to resolve the spectral indices of
the extended emission and the hotspots and test models for localised
heating (\citet{Harper01} and references therein). For example, a
decrease of $T_{\mathrm e}$  with decreasing frequency above a hot spot
would suggest a chromospheric, rather than convective origin.

The average radius \citep{Lim98} is approximately related to the
observing frequency in GHz by $R(\nu) \sim 55 \nu^{-0.5}$ au.  At a
speed of 10 km s$^{-1}$ (within the range shown by chromospheric
lines, \citealt{Gilliland96}), it would take 0.5 yr for a disturbance
to propagate 1 au from surfaces optically thick at 25 GHz to 21 GHz,
and 1 yr from the 6 GHz to the 5 GHz surface.  Monitoring at
decreasing frequencies, i.e. increasing radii, at suitable intervals,
will track the evolution of disturbances.  This will show whether we
are witnessing phenomena propagating outwards smoothly, or localized
changes behaving differently, helping to distinguish between
convection and pulsation.  We will resolve CO, other molecules and
dust using ALMA, thus measuring the composition and mass of the wind.
Clumps will be resolved on scales of a few au, within a few tens au
from the star, where chemistry is still active, overlapping with the
scales resolved in the MIR \citep{Kervella11}.

\section{Summary}
\label{sec:summary}

e-MERLIN has produced the highest-resolution images of Betelgeuse at
cm wavelengths, in July 2012.  The 5.75 GHz
photosphere is fitted by a 2-D Gaussian component, axes
(0.235$\times$0.218) arcsec$^2$ ($\sim5 R_{\star}$), flux density 1.62
mJy, with an average brightness temperature $\sim 1250$ K.
Extrapolating to 4.9 GHz shows that the flux density decline since
1981 may have levelled off; the large-scale size and brightness is
consistent with the results from \citet{Lim98}. The peak position is
close to the position predicted from the proper motion analysis by
\citet{Harper08}.   An arc is seen 0.2--0.3 arcsec to
the SW, brightness temperature $\sim 150$ K, hydrogen mass
$\sim1.8\times10^{-6}$ M$_{\odot}$.  It is in the same direction as IR
plumes seen on smaller and larger scales (\citealt{Kervella09};
\citealt{Kervella11}), a faint 4.9-GHz feature at 0.6 arcsec seen in
1996 and CO emission imaged by \citet{OGorman12}.  This suggests
persistent ejection in this direction for several centuries.

 e-MERLIN resolved two distinct peaks, 90 mas apart within the larger
stellar disc, with brightness temperatures 4000--6000 K. These are
likely to be hot spots at $\sim$5 $R_{\star}$, although it is 
possible  that cooler, low-opacity inhomogeneities in higher layers reveal the hottest
chromospheric layers.  Similar, isolated peaks
were also detected by MERLIN observations in 1992, 1995 and 1996.
Such high-sensitivity, high-resolution radio continuum observations
can resolve emission simultaneously from material at chromospheric and
photospheric temperatures, and have the potential to track proper
motions.  Future multi-frequency imaging with e-MERLIN, the VLA and
ALMA will investigate whether the signatures of chromospheric
emission, convection or pulsation shocks are seen in the radio
photosphere from 2--10 $R_{\star}$, helping to elucidate the
mechanisms which eject material from the star into its envelope.

\section{Acknowledgments} We warmly thank M. K. Argo, R. Beswick, and
the rest of the e-MERLIN team for guidance in reducing these
data. e-MERLIN is the UK radio interferometer array, operated by the
University of Manchester on behalf of STFC.  We are very grateful to
the referee for insightful comments which improved the accuracy of
this paper.  We acknowledge the use of MERLIN and VLA archival data.

\bibliographystyle{mn2e}

\bibliography{cse}

\newpage
\begin{appendix} 
\section{Data processing summary}

\subsection{Calibration} This is a summary of the steps for data
calibration.  {\small {AIPS}} \citep{Greisen94} was used for data
processing as the required capabilities are not yet fully implemented
in {\small {CASA}}.  Data were stored in multi-source format, the
time-dependent solution tables being interpolated into calibration
tables using {\small {CLCAL}}.  Separate solutions were derived for
each spectral window and hand of circular polarization.

\begin{enumerate}
\item Load data ({\small {FITLD}}), ensure that they are in
time-baseline order ({\small {MSSORT}}) and inspect/edit.  Editing
({\small {SPFLG}}, {\small {UVFLG}} and {\small {IBLED}}) was
performed at a number of stages as required.
\item Derive delay and rate corrections per scan for calibration
sources 0551+0829 and OQ208 ({\small {FRING}}).
\item Derive time-dependent phase, and then amplitude and phase
solutions for calibration sources ({\small {CALIB}}).
\item Use the solutions for the three most similar antennas (Darnhall,
Pickmere, Knockin) for a selected time interval, to establish the flux
density of 0551+0829 by comparison with OQ208 ({\small {GETJY}}),
refining the solutions by requiring a straight-line spectral index
({\small {SOUSPC}}).
  \begin{tabular}{lllll} Centre freq. (GHz)&5.4954&
5.6874&5.8154&5.9434 \\ OQ208 (Jy)&2.48&2.49& 2.50&2.51\\ 0551+0829
(Jy)&0.183& 0.190& 0.196&0.203\\
\end{tabular} \\ The flux density of OQ208 was derived from 3C286
(R. Beswick, private communication).
\item Derive channel-by-channel amplitude and phase bandpass solutions
for OQ208, averaging each half-hour scan ({\small {BPASS}}).
\item Apply the bandpass solutions using {\small {SPLAT}} and repeat
steps (ii), (iii) and (v), using the established flux density of
0551+0829 as its model.
\item Split out Betelgeuse, applying all calibration.
\item Make a preliminary, naturally-weighted image ({\small
{IMAGR}}). The star was clearly resolved, showing a bright central
ridge and a halo a few hundred mas in size, strongest to the
SW. {\small {AIPS}} fitted beam size of (0.195$\times$0.074) arcsec,
PA 143$^{\circ}$.  The major axis is more than double the true
resolution (even at declination $7^{\circ}$), because the great
sensitivity of the Lovell compared to the other telescopes creates a
bias towards shorter baselines.  All imaging used careful boxing and a
small number of iterations (50--100).
\item Use the clean components as a model for phase-only
self-calibration with a 7-min (per-scan) solution interval. These were
applied and a new image of Betelgeuse was made.  We also plotted the
visibility amplitudes and noticed some residual, slowly-changing
amplitude errors (lower amplitudes on shorter baselines, differences
between hands of polarization).  We carried out a final amplitude
self-calibration with a 1 hr solution interval, which did not affect
the source structure but improved the signal to noise ratio.

\end{enumerate}
\subsection{Imaging} All images were made in total intensity,
combining all spectral windows using a spectral index of 1.32 (using a
flat spectral index made a very small difference, much less than the
10\% total flux scale uncertainty).  There is a trade-off between
sensitivity to extended structure and resolution, in interferometry
data.  We chose two weighting schemes, based on the inherently
different sensitivities of the e-MERLIN telescopes and on empirically
established values to optimise either sensitivity or resolution
without increasing the noise more than 10\%.  The Defford telescope is
least sensitive and was given a weight of 0.1. The 75-m Lovell
telescope (providing many short baselines) and the 32-m Cambridge
telescope (providing the longest baselines) are the most sensitive at
5--6 GHz.  We maximised sensitivity to extended structure by giving
Lovell and Cambridge data weights of 10 and 2.4, respectively. We
maximised resolution by giving both the Lovell and Cambridge data a
weight of 4. All other (25-m) telescopes were given a weight of 1.
The weighting was further modified in the {\small {AIPS}} task
{\small {IMAGR}} by applying suitable interpolation into the missing
spacings; a uvtaper of 3500 k$\lambda$ with natural weighting for
optimum sensitivity and partial uniform weighting {\small {ROBUST
0.75}} for optimum resolution.

\subsection{Astrometry and photometry} 
We observed Betelgeuse at a
pointing position of J2000 05:55:10.33 +07:24:25.6.  The phase
reference ICRF2 position of 05:51:11.2293 +08:29:11.221 is accurate to
$<0.001$ arcsec and it is only $~1.\!^{\circ}5$ from Betelgeuse.
Inspection of the OQ208 raw phase corrections suggests this introduces
an uncertainty of one turn of phase in 30 min of time on the longest
baselines, corresponding to one beam in $7.\!^{\circ}5$ of arc or
0.016 arcsec in $1.\!^{\circ}5$; the uncertainties in telescope
positions introduce negligible errors.  We measured the position of
Betelgeuse for astrometric purposes by fitting a 2D Gaussian component
to the naturally-weighted image before self-calibration.

The flux density scale is derived from 3C286; we do not yet have a
full model of this source at e-MERLIN resolution and sensitivity so we
conservatively estimate a 10\% flux scale uncertainty. 0551+0829 has a
spectral index of 1.3. We note that MERLIN observations in 1995 found
a flux density of 0.128 Jy for 0551+0829 at 4.994 GHz, 20\% lower than
the flux density extrapolated from current observations.  This is
within the plausible variability of a compact QSO in 17 yr.

\end{appendix}
\label{lastpage}

\end{document}